\documentstyle[prb,aps]{revtex}

\def\avg#1{\langle #1\rangle}
\def\bx{{\bf x}}
\def\bR{{\bf R}}

\begin{document}

\twocolumn[\hsize\textwidth\columnwidth\hsize\csname @twocolumnfalse\endcsname
\widetext

\title{An ab initio path integral Monte Carlo simulation method for molecules
and clusters: application to Li$_4$ and Li$_5^+$}

\author{Ruben O. Weht~$^{\rm a,b}$, Jorge Kohanoff~$^{\rm a}$~\cite{corresp}, 
Dar\'{\i}o A. Estrin~$^{\rm a,c}$, and Charusita Chakravarty~$^{\rm a,d}$}

\address{
$^{\rm a}$ International Centre for Theoretical Physics, \\
Strada Costiera 11, I-34014 Trieste, Italy \\ 
$^{\rm b}$ Comisi\'on Nacional de Energ\' \i a At\'omica, \\
Avenida Libertador 8250, 1429, Buenos Aires, Argentina \\ 
$^{\rm c}$ Departamento de Qu\' \i mica Inorg\'anica, Anal\' \i tica y
Qu\' \i mica-F\' \i sica e INQUIMAE,\\ 
Facultad de Ciencias Exactas y Naturales, Universidad de Buenos Aires\\
Ciudad Universitaria, Pabell\'on II, 1428, Buenos Aires, Argentina \\ 
$^{\rm d}$ Department of Chemistry, Indian Institute of Technology - Delhi \\
Hauz Khas, New Delhi 110016, India 
}
\date{\today}
\maketitle

\begin{abstract}

A novel method for simulating the statistical mechanics of molecular systems
in which both nuclear and electronic degrees of freedom are treated
quantum mechanically is presented. The scheme combines a path integral 
description of the nuclear variables with a first-principles adiabatic 
description of the electronic structure. The electronic problem is solved 
for the ground state within a density functional approach, with the 
electronic orbitals expanded in a localized (Gaussian) basis set. 
The discretized path integral is computed by a Metropolis Monte Carlo 
sampling technique on the normal modes of the isomorphic ring-polymer. An 
effective short-time action correct to order $\tau^4$ is used.  
The validity and performance of the method are tested by studying two 
small Lithium clusters, namely Li$_4$ and Li$_5^+$. Structural and 
electronic properties computed within this fully quantum-mechanical scheme
are presented and compared to those obtained within the classical nuclei 
approximation. Quantum delocalization effects turn out to be significant 
as shown by the fact that quantum simulation results at 50 $K$ approximately 
correspond to those of classical simulations carried out at 150 $K$. The
scaling factor depends, however, on the specific physical property, thus 
evidencing the different character of quantum and thermal correlations. 
Tunneling turns out to be irrelevant in the temperature range investigated 
(50 $K$ to 200 $K$).

\end{abstract}

\vspace*{1.0cm}
]
\narrowtext

\section{Introduction}

\par Light atoms, such as H, He, Li or Be, cannot very often  be
treated as classical particles, particularly at low temperatures.
As temperature decreases and the thermal de Broglie wavelength
increases, the
quantum character emerges, and a description in terms of classical
coordinates and momenta breaks down. The most obvious manifestation
of the quantum character of light atoms is a large zero-point-energy 
(ZPE). A particle of mass $m$ in a harmonic potential with 
characteristic frequency $\omega$ will have a ZPE of $\hbar\omega/2$ 
and an associated spatial delocalization of 
$\Delta x=\sqrt{\hbar/m\omega}$. For instance, a proton in a typical 
bonding environment such as a H-O bond or a H$_2$ molecule, will
have a ZPE of 0.15 to 0.25 eV and 
$\Delta x$ between 0.2 and 0.3~\AA. This represents a sizeable effect 
which could be decisive in stabilizing a particular crystalline structure 
for a solid, or the ground state configuration of a molecule or a cluster. 
An even more interesting manifestation of quantum effects
is the possibility that these light nuclei can tunnel across
potential energy barriers, thus exploring classically forbidden regions 
of configuration space and giving rise to a variety of 
interesting quantum effects such as temperature-independent diffusion, 
exotic ground states, resonances in ion-surface scattering, and 
fluxional molecules. Signatures of quantum effects can also be 
observed in low-energy atomic collisions, or in proton-transfer 
reactions in the gas and condensed phases.

To date, most studies that consider the quantum character of
atomic nuclei are based on an empirical description of the interatomic 
interactions, or otherwise consist of extending and/or correcting {\it 
a posteriori} the results obtained within a classical nuclei approximation. 
Classical potentials are frequently not transferable from one environment
to another, and are ill-suited to modeling phenomena involving significant 
electronic density redistribution, as in the making and breaking 
of chemical bonds. The natural route to overcome these limitations is to 
describe the interactions at a first-principles level, i.e. by including 
explicitly the electronic component in the description of the system. 
The recent development of such schemes, which address the question of the 
interplay between electronic structure and the quantum nature of light 
nuclei, has only very recently began to be realized, thus opening a 
fascinating field with important implications in many branches of physics, 
chemistry and biology.

Since small clusters usually exhibit rich landscapes of isomeric forms 
within narrow energy bands~\cite{clusters}, they constitute good systems for 
studying the effects of quantum delocalization and associated tunneling 
behavior. Lithium clusters are particularly interesting because in addition 
to the small atomic mass, they are bound by metallic many-body interactions 
which cannot be adequately represented by means of classical interatomic 
potentials~\cite{litclus}. 

In the remainder of this Introduction we outline the methodology that 
we have developed to study this class of problems, and review the 
present understanding of Li clusters which is the test system
for our method. In Section II we introduce the theoretical framework of our 
ab initio path integral approach and discuss the approximations involved. 
Section III is devoted to the details of the practical implementation of 
the path integral Monte Carlo (PIMC) and the electronic structure methods.
In Section IV we present the validation of the electronic structure 
calculations, zero-temperature geometries and electronic properties of
Li$_4$ and Li$_5^+$ clusters. The results of our simulations 
for the classical and quantum Li$_4$ and Li$_5^+$ clusters at finite
temperatures are presented in Section V. Section VI contains our
conclusions and an assessment of the potential of this novel
simulation tool.

\subsection{Methodological aspects}

\par The goal of the present work is to introduce a novel computational 
technique for studying the statistical mechanics of isolated systems 
like clusters and molecules containing light atoms. Our approach combines 
an imaginary-time path integral description of the nuclear degrees of 
freedom~\cite{pathint} with a first-principles density functional (DFT) 
description of the electronic structure~\cite{elec}. Since the natural 
choice for investigating isolated systems is to use a localized basis set 
for the electronic orbitals, we adopt a Gaussian basis set~\cite{metecc}. 
In the present implementation the electronic structure is computed at the 
all-electron level, i.e. explicitly including core electrons. The sampling 
of the path integral is implemented using Monte Carlo (MC) techniques. 
The electronic energy is minimized for each nuclear configuration, and the 
MC rejection step is also performed using the energy calculated at the
same level of sophistication. 

Other schemes along these lines have been recently proposed by Marx and 
Parrinello~\cite{mp1}, and Cheng et al.~\cite{uzi}. At variance with our 
approach, these two methods use Molecular Dynamics (MD) for sampling the 
path integral, a plane-wave (PW) expansion for the electronic orbitals and 
treat the electron-ion interaction at the pseudopotential level. PW 
expansions with periodic boundary conditions are more appropriate for 
extended systems such as solids or liquids though they can be adequately 
modified to deal with isolated systems~\cite{mp1}. 

The evaluation of real-time path integrals, which would include the full
dynamical information, is an extremely difficult numerical task because the 
integrand is a rapid oscillatory function of the path. In the imaginary-time
framework the statistical weight is real and positive definite, so that
the integrals are well-conditioned, but the price is that the dynamics is
not directly accessible. In the absence of real-time dynamical information,
it is not particularly advantageous to sample the integral using MD in place 
of MC techniques. In particular, the MD technique requires elaborate 
thermostatting mechanisms~\cite{mp2} in order to overcome ergodicity problems 
in the sampling of the quasi-harmonic degrees of freedom that appear in the 
path integral formulation (see Section II). In the present method we propose 
a Metropolis MC sampling technique on the normal modes of the polymer, which 
has less severe ergodicity problems, and is more convenient from the
point of view of efficient evaluation of the path action (see Section 
III)~\cite{pimc}. Moreover, the MC strategy is easy to interface with 
any electronic structure code.

\subsection{Small Lithium clusters}

\par Structural and electronic properties of Li$_n$ and Li$_n^+$ clusters 
have been systematically investigated for $n=2-9$ by means of {\it ab initio} 
configuration-interaction calculations~\cite{boustani}. A key observation
was the existence of several isomers of comparable energy. Recently, ab 
initio classical MD simulations at the Hartree-Fock (HF) and non-local 
density functional (NLDF) level were carried out in order to explore the 
cluster dynamics as a function of temperature, to analyze isomerization 
reactions, and to study the melting 
transition~\cite{jellinek,reichardt,fantucci,jones}. The extent to which 
different levels of first-principles calculations are reliable for describing 
small Li clusters has been very recently discussed by Rousseau and 
Marx~\cite{rousseau}, who concluded that either {\it ab initio} MP2 
calculations, or gradient corrected NLDF provide a reasonable potential 
energy landscape, while HF and LDF are inadequate. 

This aspect is very important because the energy landscape determines 
the probability with which the cluster visits different possible geometries. 
In general, the ability to jump from one minimum to another will depend on 
the extent of both thermal and quantum fluctuations. Fig. 1 shows three 
typical situations: At low temperatures quantum delocalization is the 
dominant mechanism. If the ZPE is higher than the barrier between the two 
minima (a), or smaller than the barrier but larger than the energy difference 
between them (b) (tunneling regime), the ground state wave function will 
sample both configurational minima. If, instead, it is smaller than the 
energy difference between the two minima (c), then the ground state will be
basically unchanged from the classical one, though all structural 
distributions will be broadened by quantum delocalization. Thermal 
excitations can promote a classical-like situation like (c) to a 
thermally-assisted tunneling regime.

The above picture is valid as long as the electronic ground state is 
non-degenerate, and this is realized for clusters with closed electronic 
shells. Open-shell clusters undergo Jahn-Teller distortions, and are
characterized by degenerate ground states and {\it pseudo-rotations}.
An adequate treatment of this situation would require the introduction of 
the concepts of conical intersections and geometric phases~\cite{berry}. 
Though these phenomena are of great interest, and have indeed been detected 
in Li$_3$~\cite{li3} and Li$_5$~\cite{li5}, we have preferred to avoid this 
extra complication at this stage in the development of our method.

Quantum nuclear effects in Li clusters were first addressed by Ballone and 
Milani~\cite{ballone}, who studied magic-number clusters (20, 40 and 92
atoms) by means of PIMC simulations using a simple, jellium model 
description of the electronic component. Their most interesting observation 
was the existence of a large number of isomers differing only in the 
location of the outermost atoms, such that 
quantum tunneling amongst these isomers led to a fluid-like behavior.
A very recent ab initio path integral MD study of Li$_8$ and Li$_{20}$ 
clusters by Rousseau and Marx~\cite{marx} shows that this picture does 
not hold when the description of the electronic component is improved, 
thus demonstrating the necessity of going beyond the jellium-type 
description.

For our study we have chosen the following two examples: Li$_4$ has a 
single isomer and is well described by a quasi-classical picture. Li$_5^+$,
has two isomers at an energy difference of about 150 $K$, and two other 
isomers at higher energies (see section 5). The zero-point-energy is of 
the order of 100 $K$, so that Li$_5^+$ could constitute a good candidate 
for exhibiting significant quantum effects in terms of the sampling of
configurational minima. The results presented in section 5 will show 
that, in spite of this, Li$_5^+$ actually behaves in a very similar way
to Li$_4$.

\section{The ab initio path integral partition function}

\par The statistical mechanics of quantum many-body systems can be 
formulated in terms of the two-point density matrix, or imaginary-time 
propagator:

\begin{equation}
\rho({\bf R},{\bf R}',\beta)=<{\bf R}|\exp(-\beta\hat{\cal H})|{\bf R}'> ,
\end{equation}
 
\noindent whose trace is the partition function ${\cal Z}$. $\beta=1/k_BT$ 
is the inverse temperature. 
The path integral representation of the density matrix is given by

\begin{equation}
\rho(\bR,\bR ',\beta)=\int{\cal D}[\bR(u)]~{\rm e}^{-S[\bR(u)]} ,
\end{equation}

\noindent where $\bR(u)$ represents the configuration of an $N$-body 
system as a function of imaginary time $u$. The range of $u$ is from 0 
to $\beta\hbar$ and the paths considered are restricted to those beginning 
at $\bR(0)=\bR$ and ending at $\bR(\beta\hbar)=\bR '$. The partition 
function can be similarly expressed as a
path integral with contributions from all possible cyclic paths for which 
$\bR(0)=\bR(\beta\hbar)$. ${\cal D}[\bR(u)]$ represents the differential 
element for all paths. The Euclidean action, $S[\bR(u)]$, associated with 
a path is defined as

\begin{equation}
S[\bR(u)]=\frac{1}{\hbar}\int_0^{\beta\hbar}\Bigl(~\frac{m}{2}~
\Bigl[\frac{d\bR}{du}\Bigr]^2~+~V\Bigl[\bR(u)\Bigr]~\Bigr)du .
\end{equation}

The first term corresponds to the kinetic energy contribution to the action,
with $m$ the mass of the particles. The generalization to heterogeneous
systems, i.e. composed of two or more species of different masses, is 
straightforward. 

In order to devise a feasible computational scheme,
the path integral is typically discretized by representing the cyclic
paths as a finite set of $3N$-dimensional configurations, $\bR_i$,
at equispaced points in imaginary time between $0$ and $\beta\hbar$.
The degree of discretisation is referred to as the Trotter number, $P$.
The short-time or high temperature propagator, 
$\rho (\bR_i ,\bR_{i+1} ;\beta /P)$, can be evaluated semi-classically
at different levels of approximation. The contribution of the kinetic 
energy term to the short-time action is written in terms of a first-order 
finite difference between configurations on adjacent time slices, while 
the short-time integral of the potential 
energy together with higher-order corrections to the kinetic energy, 
is replaced with an effective, quantum-corrected potential 
$V_{eff}(\bR)$. The sum of the two terms is referred to as the {\it 
effective action}. Therefore, the expression for the partition function of 
$N$ interacting, distinguishable quantum particles with Trotter number $P$ 
is given by:

\begin{eqnarray}
{\cal Z}_{NP}=\left(\frac{mP}{2\pi\hbar^2\beta}\right)^{3NP/2}
\int~\Biggl(\prod_{i=1}^{P}d\bR_i\Biggr)\times
\hspace{2cm} \nonumber \\
~\exp\left(-
\frac{mP}{2\hbar^2\beta}\sum_{i=1}^{P}({\bf R}_i-{\bf R}_{i+1})^2-
\frac{\beta}{P}\sum_{i=1}^{P}V_{eff}({\bf R}_i)\right) .
\label{eq:part}
\end{eqnarray}

According to the level of approximation of the effective action, the 
number of slices $P$ needed to achieve convergence in the partition 
function can be small enough that the problem is tractable, or large 
enough that the evaluation of the multidimensional integral becomes a 
hopeless task. It is therefore important to use the best possible 
effective action compatible with the computational complexity involved in 
its calculation. The simplest one, or {\it primitive approximation}, 
replaces the effective potential by the bare potential, which is equivalent 
to an end-point approximation for the short-time integral:

\begin{equation}
\frac{1}{\hbar}\int_0^{\tau\hbar}V\Bigl[\bR(u)\Bigr]~du\approx
\tau\Biggl[\frac{V({\bf R})+V({\bf R}')}{2}\Biggr] ,
\end{equation}

\noindent and is correct only to order $\tau^2$. At the other extreme,
the pair-action approximation provides a very accurate technique when
the full, many-body potential can be reasonably approximated with 
a sum of pair potentials. This scheme has been very effectively exploited 
to investigate the properties of liquid and superfluid He down to 
temperatures of about 1 $K$~\cite{pollock}, a task that would not have 
been possible using the primitive action.

As mentioned in the introduction, classical interaction potentials are 
computationally fast, but very often unreliable. Realistic interaction
potentials can instead be obtained from more expensive first-principles 
techniques. In these latter, the electronic degrees of freedom are explicitly 
included in the Hamiltonian description of the system: 

\begin{equation}
\hat{\cal H}({\bf R},{\bf r})~=~\hat T_{\rm n}+\hat T_{\rm e}+
\hat V_{\rm ee}({\bf r})+\hat V_{\rm en}({\bf r},{\bf R})
+\hat V_{\rm nn}({\bf R}) ,
\end{equation}

\noindent where ${\bf r}$ and ${\bf R}$ are the electronic and nuclear 
coordinates, $\hat T$ and $\hat V$ stand for kinetic and potential 
operators, while subscripts {\bf e} and {\bf n} indicate electronic and 
nuclear components, respectively. The path integral representation for the 
partition function could then be developed using the coordinate basis 
for both the electrons and the nuclei~\cite{carlo}. 

However, standard electronic structure calculations are carried out in a 
wave function representation by resorting to the adiabatic separation of 
nuclear and electronic motion. It is therefore more convenient to expand the 
electronic component in the adiabatic basis set where electronic wave functions
$|\phi_{\alpha}>$ and total energies $E_{\alpha}({\bf R})$ are obtained by 
diagonalising the electronic Hamiltonian $\hat T_{\rm e}+\hat V_{\rm ee}
({\bf r})+\hat V_{\rm en}({\bf r},{\bf R})+\hat V_{\rm nn} ({\bf R})$. 
If $\tau =\beta /P$, then the discretized partition function reads:

\begin{equation}
{\cal Z}_P=\sum_{\alpha_1}\cdots\sum_{\alpha_P}~\int\cdots\int~
\prod_{i=1}^{P}~\Biggl(\rho_{\alpha_i,\alpha_{i+1}}({\bf R}_i,
{\bf R}_{i+1},\tau)~d\bR_i\Biggr) ,
\end{equation}

\noindent where

\begin{equation}
\rho_{\alpha,\gamma}({\bf R},{\bf R}',\tau)=
<{\bf R}|<\phi_{\alpha}|~\exp(-\tau\hat{\cal H})~
|\phi_{\gamma}>|{\bf R}'> .
\end{equation}

\noindent Then, in the spirit of expression (\ref{eq:part}), the 
short-time propagator can be written as:

\begin{eqnarray}
\rho_{\alpha,\gamma}({\bf R},{\bf R}',\tau)=
<{\bf R}|<\phi_{\alpha}|~\exp(-\tau\hat T_{\rm n})~
|\phi_{\gamma}>|{\bf R}'> \times
\hspace{0cm} \nonumber \\
\exp\left(-\frac{\tau}{2}\left[E^{eff}_{\alpha}({\bf R})+
E^{eff}_{\gamma}({\bf R}')\right]\right) ,
\label{eq:prop}
\end{eqnarray}

\noindent where $E^{eff}_{\alpha}({\bf R})$ is an effective potential which
derives directly from the electronic structure. In the primitive 
approximation, which is what has been used in other ab initio path integral 
methods~\cite{mp1,uzi}, the effective potential is simply the total energy 
corresponding to adiabatic state $\alpha$ of the electronic Hamiltonian, 
i.e. $E_{\alpha}({\bf R})$. 

The nuclear kinetic energy operator can give rise to non-adiabatic coupling 
matrix elements between adiabatic eigenstates. If these are negligible, but 
more than one Born-Oppenheimer (BO) surface is occupied due to thermal 
excitations, then the partition function will split into independent 
manifolds indexed by the BO electronic eigenstate. In the absence of 
degeneracies in the ground electronic state, the energy differences 
between electronic eigenstates are typically orders of magnitude larger 
than reasonable thermal kinetic energies. Consequently only the ground 
electronic state contributes to the partition function and the 
electrons only enter at the level of replacing the total potential energy 
with the ground state first-principles effective potential 
$E^{eff}_0({\bf R})$ in equation (\ref{eq:part}):

\begin{eqnarray}
{\cal Z}_{NP}=\left(\frac{mP}{2\pi\hbar^2\beta}\right)^{3NP/2}
\int~\Biggl(\prod_{i=1}^{P}d\bR_i\Biggr) \times
\hspace{2cm} \nonumber \\
~\exp\left(-
\frac{mP}{2\hbar^2\beta}\sum_{i=1}^{P}({\bf R}_i-{\bf R}_{i+1})^2-
\frac{\beta}{P}\sum_{i=1}^{P}E^{eff}_0({\bf R_i})\right) .
\end{eqnarray}

\noindent This is the partition function that will be evaluated using 
Monte Carlo techniques. The expression for ${\cal Z}_{NP}$ can be 
interpreted as the partition function of $N$ classical polymers, each of 
$P$ monomeric units or beads, with adjacent beads linked by harmonic 
springs with force constant $mP/\beta\hbar^2$. Beads on two separate 
cyclic polymers are coupled by the interaction potential only if they 
lie on the same time slice.

Let us remark that the computation of excited electronic states is an 
open issue in density functional theory, and it is known that the usual
approximations to exchange and correlation, like the LDA and even NLDF
do not provide reliable excitation energies. Excited states could be 
calculated properly using very high-level {\it ab initio} methods. However, 
since we have developed a methodology in which the electronic degrees of 
freedom are dealt within DFT, we are for the time being not in a position 
to incorporate non-adiabatic couplings and excited electronic manifolds.

\section{Practical implementation}

\subsection{Path integral Monte Carlo}

\subsubsection{Effective action} 

\par We use a discretized time representation in which a path is 
described as a set of configurations, $\{\bR_i\}, i=1,\cdots P$, at 
$P$-equispaced points in imaginary time. The effective short-time propagator 
for two adjacent points along the path has been evaluated to fourth-order 
accuracy in $\tau=\beta/P$, so that the first-principles 
effective potential reads\cite{ponjas}:

\begin{equation}
E^{eff}_0(\bR_i)=E_0(\bR_i)+\left(\frac{\beta^2\hbar^2}{24mP^2}\right)~
\sum_{j=1}^N\left(\frac{\partial E_0(\bR_i)}{\partial \bx_{ij}}\right)^2,
\end{equation}

\noindent where $\bx_{ij}$ is the 3-dimensional vector of the coordinates 
of particle $j$ in slice $i$, such that $\bR_i=(\bx_{i1},\bx_{i2},\cdots,
\bx_{iN})$.  This form of the effective action leads to an error
of the order of $P(\beta /P)^5$ in the partition function and allows us
to significantly reduce the Trotter number required for convergence as
compared to the primitive action (see Section V).

The quantum correction to the potential requires the evaluation of the
first-principles forces ${\bf F}_j=-\partial E_0/\partial\bx_j$ in the 
ground electronic state. The cost of this operation is of the same order
of magnitude as the rest of the electronic structure calculation, at least
in most {\it ab initio} or density functional schemes. In contrast, second 
and higher-order derivatives are sufficiently expensive to evaluate that 
the computational advantages of a more accurate short-time action are lost.

\subsubsection{Normal modes sampling}

\par The above expression for the partition function can be directly
used to set up a Metropolis MC simulation scheme by assigning the 
appropriate Boltzmann weight to each configuration of the $N\times 
P$-dimensional isomorphic classical system. However, as quantum effects 
increase, the 
degree of discretisation must be increased to maintain accuracy. Since the 
harmonic force constant between adjacent beads on the quantum polymer is 
$mP/\beta\hbar^2$, increasing the Trotter index results in increasingly
stiff harmonic links and the computational problem of ensuring the
ergodicity of the Metropolis walk becomes intractable. An intuitively 
appealing and computationally simple way for circumventing this 
difficulty comes from considering the normal modes of the quantum polymer
\cite{normal}. In the absence of an interaction potential, all Cartesian 
degrees of freedom of the system are decoupled. For a single degree of 
freedom the harmonic intra-polymer potential is given by
$V_p=(mP/2\beta\hbar^2)\sum_{l=1}^P(x_l-x_{l+1})^2$. Diagonalization of 
the second derivative matrix of this potential leads to the normal 
coordinates, $\{Q_k\},k=1,\cdots,P$,

\begin{equation}
Q_k = (1/\sqrt{P})\sum_{l=1}^Px_l\exp (2\pi ikl/P) .
\end{equation}

In the normal mode representation, the kinetic energy contribution 
to the path action is

\begin{equation}
\int_0^{\beta\hbar}{m\over 2}\Bigl({d\,x\over du}\Bigr)^2\, du
={2mP\over \beta\hbar^2}\sum_{k=1}^P\vert Q_k\vert^2\sin^2(\pi k/P) .
\end{equation}

\noindent The zero-frequency mode ($k=P$)
corresponds to motion of the center-of-mass of the polymer and makes no
contribution to the kinetic energy. All other normal modes would be Gaussian 
distributed with variance 
$\sigma_k^2=\beta\hbar^2 /4mP\sin^2(\pi k/P)$ if they corresponded to 
free particles. The potential energy term couples these normal modes and 
cause distortions from the free-particle distribution. The low-frequency 
modes correspond to large, collective motions of all beads of the polymer, 
while the high-frequency modes cause small, local path fluctuations. The 
normal modes are then used as Metropolis variables and the displacements 
scaled according to the Gaussian dispersions associated with each normal mode.

\subsubsection{Observables}

\par The canonical ensemble average of an observable $O$ is given by

\begin{equation}
\langle O \rangle = Tr\{\hat\rho \hat O\}/Tr\{\hat\rho\} ,
\end{equation}

\noindent where $\hat O$ is the corresponding quantum mechanical operator.
If the operator $\hat O$ is diagonal in the coordinate representation, then

\begin{equation}
\avg{O} =\int~d\bR~O(\bR )~\rho (\bR ,\bR ;\beta ) .
\end{equation}

Our MC strategy samples configurations {\bf R} with probability proportional 
to $\rho (\bR ,\bR ;\beta )$, such that equilibrium averages can be readily 
estimated via discrete summations.
Dynamical variables that can be related to the partition function are
also straightforward to obtain using thermodynamic estimators.
With our choice of the short-time action, thermodynamic
estimators of the total energy $\avg E$, the kinetic
energy $\avg K$ and the potential energy $\avg V$ are given by

\begin{eqnarray}
\avg V &=& \avg{U_2} + 2\avg{U_c}\\ \nonumber \\
\avg K &=& \avg{U_1} +\avg{U_c}\\ \nonumber \\
\avg E &=& \avg{U_1}+\avg{U_2} + 3\avg{U_c} ,
\end{eqnarray}

\noindent where 

\begin{eqnarray}
U_1 &=& {3NP\over 2\beta}~-~\sum_{i=1}^{P}{mP(\bR_i -\bR_{i+1})^2\over
2\beta^2\hbar^2}\\ \nonumber \\
U_2 &=& \frac{1}{P}~\sum_{i=1}^{P}E_0(\bR_i)\\ \nonumber \\
U_c &=& \frac{1}{P}~\frac{\beta^2\hbar^2}{24mP^2}~\sum_{i=1}^{P}~
\sum_{j=1}^N~\Bigl(\frac{\partial E_0(\bR_i)}{\partial\bx_{ij}}\Bigr)^2 .
\end{eqnarray}

\subsection{Electronic structure calculations}

\par The calculation of the electronic energies is carried out within the 
framework of DFT. For a given nuclear configuration, Kohn-Sham 
single-particle equations~\cite{ks} are solved self-consistently for the 
electronic density, and the total energy and forces are computed 
accordingly. Kohn-Sham orbitals are expanded in a Gaussian basis set.

The electronic density is also expanded in an additional Gaussian basis
set~\cite{dunlap}. The coefficients for the fit of the electronic density 
are computed by minimizing the error in the Coulomb repulsion energy. The 
use of this procedure results in an important speedup, since the cost
of evaluating matrix elements reduces from $O(N^{4})$ to $O(N^{2} M)$ 
($N$ is the number of functions in the orbital set, and $M$ the number of 
functions in the auxiliary set, typically of size comparable to $N$). 

Matrix elements of the exchange-correlation potential are calculated
by a numerical integration scheme based on grids and quadratures 
proposed by Becke~\cite{becke}. During the self-consistency cycle, 
the integration is performed on a set of coarse atom-centered, spherical
grids. At the end of the self-consistent procedure, the exchange-correlation 
energy is evaluated using an augmented, finer grid. This strategy of 
combining coarse and fine grids results in a considerable gain in 
computational efficiency, which is very important because this part is 
one of the main bottlenecks of the calculation.

The exchange-correlation term is described a gradient corrected NLDF
level. Correlation is given by the parameterization
of the homogeneous electron gas of Vosko et al.~\cite{vwn} supplemented
with the gradient corrections proposed by Perdew~\cite{perdew}. Gradient 
corrections to the exchange term are taken from Becke~\cite{becke2}.

The first derivatives of the energy with respect to the nuclear coordinates,
required by the fourth-order effective action, are evaluated by taking 
analytical derivatives of one-electron and Coulomb terms, while the
exchange-correlation contribution is obtained by numerical 
integration~\cite{metecc}.

\section{Ground state properties of classical Li$_4$ and Li$_{5}^{+}$}

\subsection{Validation of the basis set and optimized geometries}

\par We have analyzed five different basis sets for Li$_{4}$ and 
Li$_{5}$$^{+}$. The first one (labeled 1) is the standard 3-21G 
basis~\cite{binkley}. The second set (labeled 2) is a double zeta plus 
polarization basis set, optimized for DFT calcultions~\cite{godbout}. 
The third one (labeled 3) is the standard 6-311G basis~\cite{krishnan}, 
the fourth set (labeled 4) is the 6-311G set augmented with a polarization 
function (6-311G*), and finally the fifth set (labeled 5), consists of a 
large uncontracted basis set (13s/9p/1d), proposed by 
Dunning~\cite{dunning}. The calculations performed with basis sets 
optimized for standard ab-initio calculations (labeled 1, 3, 4, and 5) 
have been carried out using an uncontracted auxiliary basis set 
with a scheme (7s/3p/3d), as proposed in Ref.~\onlinecite{godbout} 
and~\onlinecite{boustani}. The calculations performed with basis set 2 
have been carried out using an auxiliary basis set proposed in 
Ref.~\onlinecite{godbout}, 
with a scheme (7s/2p/1d). Full geometry optimizations without symmetry
constraints have been performed in all cases.

In agreement with previous work~\cite{boustani,rousseau}, only one stable 
minimum with a rhombus geometry has been found for Li$_{4}$ (see Fig. 2),
while for Li$_{5}$$^{+}$ we found four stable local minima (also shown in 
Fig. 2). The highest energy isomer, i.e. isomer I, consists of two triangles 
which lie on the same plane, joined by a shared central atom. The second 
isomer, i.e. isomer II, is similar to isomer I, but now the triangles lie 
on perpendicular planes. The third isomer, i.e. isomer III, has C$_2$ 
symmetry, and can be described as an isosceles triangle plus a dimer. 
The dimer is located perpendicularly to the plane of the triangle, close 
to its shortest side. The fourth isomer, i.e. isomer IV, has a trigonal 
bipyramidal structure. It should be pointed out that neither isomer II nor 
isomer III have been reported in earlier works, probably because of
symmetry constraints used during the geometry optimization procedure.
The basis set dependence of the binding energies is shown in Figure 3. 
Relevant structural parameters for all basis sets considered here are given 
in Table I. Our results for basis set 4 agree with those reported for the 
same functional and basis set in Ref.~\onlinecite{rousseau} for Li$_{4}$ 
and  for isomers I and IV of Li$_5^+$.

The data in Figure 3 and Table I show that calculations performed using 
basis set 2 yielded results that deviate considerably from the ones
obtained using the very large basis set 5, which can be considered as
almost converged. Even if the errors in computed bond lengths and
binding energies are not too large (about 5 \% for bond lenghts and
5-10 \% for binding energies), calculations carried out using basis set 2 
fail in reproducing the energy sequence of isomers of Li$_{5}$$^{+}$. 
This can be ascribed to the fact that the description of the $p$-shell 
using this set is rather poor, since it contains only one function.
On the other hand, calculations performed with basis sets 1, 3 and 4 
yield the same energy ordering for isomers of Li$_{5}$$^{+}$ and structural 
results within 2-3\% from the almost converged, basis set 5, values.

Basis set superposition error (BSSE)~\cite{boys} calculations for Li$_{4}$ 
yield 14.36, 1.00, 0.96, 2.34, and 0.13 kJ/mol, for basis sets 1, 2, 3, 
4, and 5, respectively. It is clear that in order to reduce BSSE, and
obtain meaningful interaction energies, a better description of the 
$p$-shell than the one provided by the small basis set 1 is required.

\begin{table}
 
\caption{Selected geometrical parameters of Li$_{4}$ and Li$_{5}$$^{+}$
(in~\AA). Atoms are labeled as in Figure 2.\\}
\vspace{0.3truecm}
 
\begin{tabular}{llccccc}
 
 System &  &  Set 1 & Set 2 & Set 3 & Set 4 & Set 5 \\ \hline\hline
 Li$_{4}$  &  d$_{12}$ & 2.658 & 2.678 & 2.638 & 2.625 & 2.622 \\
          &  d$_{13}$ & 3.068 & 3.117 & 3.050 & 3.042 & 3.039 \\ \hline
 Li$_{5}^+$  &  d$_{23}$ & 2.879 & 2.889 & 2.853 & 2.853 & 2.853 \\
 isomer I  &  d$_{12}$ & 3.114 & 3.144 & 3.105 & 3.105 & 3.104 \\ \hline
 Li$_{5}^+$  &  d$_{23}$ & 2.888 & 2.905 & 2.868 & 2.860 & 2.854 \\
 isomer II &  d$_{12}$ & 3.099 & 3.155 & 3.101 & 3.090 & 3.095 \\ \hline
 Li$_{5}^+$   &  d$_{14}$ & 2.713  & 2.735 & 2.695 & 2.680 & 2.672 \\
 isomer III &  d$_{23}$ & 2.851  & 2.857 & 2.842 & 2.823 & 2.825 \\
           &  d$_{15}$ & 3.035  & 3.053 & 3.012 & 3.012 & 3.004 \\ \hline
           &  d$_{34}$ & 3.353  & 3.485 & 3.360 & 3.345 & 3.352 \\
 Li$_{5}^+$  &  d$_{12}$ & 2.772 & 2.800 & 2.754 & 2.734 & 2.729 \\
 isomer IV &  d$_{14}$ & 3.207 & 3.250 & 3.196 & 3.186 & 3.184 \\
\end{tabular}
\end{table}

In view of these results, the intermediate basis set 3 (6-311G), which 
yields results close to the ones obtained with the larger sets is chosen
to perform the electronic structure calculations required in the MC
simulations. The relative energies with respect to the most stable isomer,
isomer IV, for isomers I, II, and III, are 17.58, 15.16, and 9.67 kJ/mol,
respectively, for calculations using basis set 3, compared with
19.38, 17.08, and 11.01 kJ/mol, respectively, using the large set 5.

\subsection{Electronic properties: dipole moments, Mulliken population
charges and eigenvalues} 

\par The Li$_{4}$ cluster is non-polar (vanishing dipole moment) for
symmetry reasons. Li$_{5}$$^{+}$ is a charged system, so its dipole moment 
depends on the choice of the origin. However, it is customary to evaluate 
the dipole moment using the center of charge as the origin, and in that case 
it provides a useful indicator of the asymmetry of the charge distribution, 
and could also be experimentally relevant. Isomers I, and II are non polar, 
isomer IV is only slightly polar, but isomer III is considerably polar. The 
dipole moments of isomers III and IV, computed using basis set 3, are 1.163 
$D$ and 0.017 $D$, respectively. 
Mulliken population charges~\cite{szabo} are also useful indicators of the 
charge distribution. Results obtained with basis set 3 for Li$_{4}$ and the 
four isomers of Li$_{5}$$^{+}$ are shown in Table II. Significant 
differences are observed between different isomers, even between isomers I 
and II which are very similar both, geometrically and energetically.
Both quantities provide useful indicators of isomerization during the MC 
simulations in the Li$_{5}$$^{+}$ case. 

In addition, we present the two highest occupied and two lowest unoccupied 
molecular orbital energies for Li$_{4}$ and the four isomers of 
Li$_{5}$$^{+}$ in Table II. The HOMO-LUMO gap is quite large in all
cases, with values around 1 eV. Therefore, it is unlikely that either
thermal or quantum fluctuations will contribute to its closure.

\vspace{0.3truecm}
\begin{table}
 
\caption{Mulliken populations and orbital energies of Li$_{4}$
and Li$_5^+$ computed with basis set 3.
(in~\AA~and eV). Atoms are labeled as in Figure 2. The two lowest
energy unocuppied and two highest energy ocuppied orbital energies
are given.}
\vspace{0.3truecm}
 
\begin{tabular}{lccccc}
  &  Li$_{4}$ & Li$_5^+$ (I) & Li$_5^+$ (II) & Li$_5^+$ (III) & Li$_5^+$ (IV)\\
\hline\hline
q$_{1}$ &  0.2067  & 0.5403  & 0.0419  & 0.2250 & 0.3959 \\
q$_{2}$ &  0.2067  & 0.1149  & 0.2395  & 0.1808 & 0.3603 \\
q$_{3}$ &  -0.2067  & 0.1149  & 0.2395  & 0.1808 & 0.3603\\
q$_{4}$ &  -0.2067  & 0.1149  & 0.2395  & 0.2250 & -0.0583 \\
q$_{5}$ &  -  & 0.1149  & 0.2395  & 0.1878 & -0.0583 \\ \hline
$\epsilon_{N-1}$ & -3.9334   & -7.2535  & -7.2355   & -7.8535  & -8.1536  \\
$\epsilon_{N}$   & -2.8986   & -6.6867  & -6.6703   & -6.5560  & -6.6603  \\
$\epsilon_{N+1}$ & -2.0169   & -5.3772  & -5.0774   & -5.4975  & -5.7065  \\
$\epsilon_{N+2}$ & -1.6806   & -5.0904  & -5.0774   & -5.4562  & -5.7002  \\
\end{tabular}
\end{table}

\section{Results of the PIMC-DFT simulations}

\subsection{Sampling strategy and convergence of the path integral with 
the degree of discretisation}

We used a simple Metropolis algorithm for the PIMC simulations with 
each trial move consisting of an attempt to move all the normal modes 
associated with all the particles. Two different step-sizes were 
used: $\delta_c$ and $\delta_s$. The maximum displacement of the 
center-of-mass was set by $\delta_{c}$, and that of the normal modes 
of order $k$ -- associated with a length scale $\sigma_k$ -- by 
$\sigma_k\times\delta_s$. We have analyzed the possibility of
introducing an additional convergence parameter $k^*$, such that modes 
with $k< k^*$ or $k> P-k^*$ are moved with a relatively small step size, 
$\delta_s\times\sigma_k$, while those associated with small 
length scale fluctuations are moved by amounts proportional to 
$\delta_l\times\sigma_k$ ($\delta_l>\delta_s$). However, it turned out 
that, in this particular case, a single step size $\delta_s$ for
all values of $k$ was efficient enough. This is often not the case when a
large number of Trotter slices is used. The various
parameters were adjusted to keep the overall acceptance ratio around 
50\% though occasional runs were used with acceptance ratios between 
40\% and 60\%. The same displacement parameter for the center-of-mass 
$\delta_c$ was used for both classical and quantum simulations. 

Simulations on the Li$_2$ dimer using a classical potential fitted to 
first-principles calculations were used to check the convergence of 
various properties with the degree of discretisation. 

Table III gives the PIMC results for the expectation
value of the potential and kinetic energies using the primitive
action and the fourth-order corrected form of the action. It can be
observed that convergence to within the statistical error bars occurs with
a Trotter number of just 4 when using the fourth-order correction,
in contrast to 16 when using the primitive action. Errors in
$\avg V$ are an order of magnitude less than those in $\avg K$.

\vspace{0.5truecm}
\begin{table}
 
\caption{Convergence data for Li$_2$ at 100 $K$. Number of MC configurations
is 5.12 million with acceptance ratios between 0.4 and 0.7. Error bars
are given in brackets. First two columns correspond to results
using the primitive approximation and the last two to results obtained using
the fourth order correction to the effective action. Energies are expressed in
$K$.
}
\vspace{0.3truecm}
 
\begin{tabular}{ccccc}
m  & $\avg V$     & $\avg K$    & $\avg V$      & $\avg K$     \\
\hline\hline
1  & -9521.78 (0.12) &  300.0 (0.0) & -9485.82 (0.14) & 335.7 (0.1)\\
2  & -9491.74 (0.15) &  330.3 (0.5) & -9454.41 (0.15) & 366.2 (0.7)\\
4  & -9465.76 (0.26) &  355.4 (1.0) & -9446.69 (0.31) & 372.8 (0.9)\\
8  & -9453.39 (0.45) &  366.5 (1.5) & -9447.10 (0.32) & 371.8 (2.4)\\
16 & -9449.79 (0.40) &  373.2 (7.6) & -9447.50 (0.44) & 371.1 (7.0)\\
\end{tabular}
\end{table}

This is typical of the relatively facile convergence of expectation
values of operators diagonal in the coordinates as opposed to those 
than must be evaluated using thermodynamic estimators. In our experience,
structural quantities such as pair distribution functions converge even 
faster than the potential energy. The increase in the error bars
-- at constant number of MC configurations -- as the Trotter number is 
increased is also typical of PIMC simulations. Based on our tests with 
Li$_2$ we used a Trotter number of 4 at 100K and 8 at 50K for the larger 
clusters.

We have performed a series of classical and quantum Monte Carlo
simulations for Li$_4$ and Li$_5$$^+$ clusters, with temperatures
ranging from 50 $K$ to 200 $K$. Each of them consisted of 10000-15000 Monte
Carlo steps, preceeded by around 1000 steps of thermalization. 
We stored atomic coordinates, energies, eigenvalues, 
Mulliken population charges and dipole moment for each MC step, for later 
analysis. 

In the following we will concentrate only in structural parameters, 
one-electron eigenvalues, and dipole moments, leaving aside other 
thermodynamical properties which would need longer simulations to 
reduce the statistical error bars to useful values.

\subsection{Results for Li$_4$ and Li$_5^+$}

\par As mentioned in Section IV, Li$_4$ has a single, deep minimum at 
the $^1$A singlet state, in the form of a planar rhombus. Due to this fact 
the cluster is very rigid and thermal and/or quantum effects 
basically sample the PES around the minimum.
The Li$_5^+$ cluster constitutes a somewhat richer example due to the 
existence of several isomers. In particular, it is interesting to
analyze the possibility of thermal activation and quantum 
tunneling between different regions of configuration space. In order
to explore different regions of the PES, classical simulations were 
started from three different isomers, namely the ground state (isomer IV), 
and two higher energy configurations (isomers I and III). After a few 
thousand MC steps it was observed that the second and third simulations 
were attracted towards the ground state basin, showing that our MC 
strategy is quite effective in equilibrating and exploring configuration 
space, possibly because interconversion barriers were low.
Based on this, quantum simulations were started from the ground state 
(isomer IV) and, in order to facilitate the detection of possible 
tunneling behavior, also from the first excited isomer (isomer III). Again,
the isomerization towards the ground state was observed in this latter;
the polymer moved as a whole, without showing any signature of tunneling.
Let us mention that none of the higher-energy isomers appeared again 
during the simulation, although structures slightly reminiscent of isomer 
III (the closest in energy to the ground state) were observed. In other
words, Li$_5^+$ appears to be unable to sample metastable regions of 
configurational space out from the ground state.

Figure 4(a) shows the radial distribution function $g(r)$ of Li$_4$ in the 
classical case, and for different temperatures. It can be observed that the
peaks are approximately centered at the optimized zero-temperature 
distances. Temperature effects consist basically of broadening the peaks; 
the first of them, corresponding to the first neighbour shell, almost 
disappears above 200 $K$. In Figure 4(b) we show the effects of the quantum 
nature of the nuclei by comparing simulations performed at 50 $K$ using
the classical and quantum schemes. It can be observed that quantum effects 
generate a pronounced broadening of the peaks, thus demonstrating the
importance of their inclusion.

In Figure 5 we show the radial distribution functions $g(r)$ for 
Li$_5^+$ in the classical (a) and quantum (b) cases. The main panels 
contain the distribution averaged over all the 5 particles. Two groups
of atoms can be identified: a first one composed of three atoms more
strongly bound, which form the central triangle of the bipyramidal
structure IV, and a second one composed by the two external atoms. 
The upper inset shows the partial $g(r)$ corresponding to these two 
groups in the classical case at 100 $K$, and the lower inset in the
quantum case at 50 $K$. No qualitative
difference with Li$_4$ can be observed. 

The trends discussed above also hold for the electronic properties, i.e. 
the distribution of one-particle eigenvalues. As can be observed in Figure 
6(a), the HOMO and LUMO eigenvalue distributions for Li$_5^+$ exhibit 
significant 
broadenings upon temperature increase. It is to be remarked that the widths 
are different for different eigenvalues, a fact that could be reflected 
in the temperature dependence of the optical photoabsorption
spectrum~\cite{rubio}. Similar broadenings can be observed in the lower 
panel, corresponding to the quantum and classical simulations for Li$_5^+$ 
performed at 50 $K$. As advanced above, the minimum HOMO-LUMO distance
never becomes smaller than about 0.5 eV, so that quantum effects cannot
promote an eigenvalue crossing which would result in a major modification
of the electronic properties. 

Another important quantity is the electric dipole moment, because of its
experimental relevance. For Li$_4$ the dipole moment vanishes at zero 
temperature due to symmetry considerations. However, at finite temperature
the cluster samples regions of the PES characterized
by a finite dipole moment, such that the mean value is non-vanishing.
A more pronounced effect of the same type can be observed in the
quantum simulations. The averages computed using the classical and quantum 
schemes at 50 $K$ are (0.17 $\pm$ 0.07) and (0.29 $\pm$ 0.15) D, for Li$_4$ 
and (0.14 $\pm$ 0.07) and (0.25 $\pm$ 0.12) D for Li$_5^+$, respectively.
The quantum dipole moment averages and distributions obtained at 50 $K$ 
are similar to those obtained classically at about 100 $K$.

It is interesting to note that where structural properties are
concerned, the overall effect of considering quantum nuclei is 
qualitatively similar to the effect of increasing temperature in the 
classical simulations. For the closed-shell Li clusters considered in
this work the classical temperature equivalent to the quantum system 
at 50 $K$ is around 150 $K$. However, the dipole moment and eigenvalue 
distributions obtained using the quantum mechanical scheme at 50 $K$ 
are qualitatively similar to those obtained within the classical scheme 
at about 100 $K$. This points out an 
important difference between the two types of correlations involved in the
statistical mechanics of quantum systems, namely coherent quantum 
fluctuations as opposed to incoherent thermal fluctuations. These
appear to behave in different ways according to the physical properties
under consideration.

Further characterization of the wave packet behaviour of the nuclei is
provided by the imaginary-time correlation function, 
$R^2(t-t')=<|r(t)-r(t')|^2>$, where $0 < \tau=t-t' < \beta \hbar$. The
value at $\tau = \beta \hbar /2$ is particularly important because it 
gives an estimation of the quantum delocalization of the nuclei. The
delocalization length, or gyration radius, is about 0.15~\AA~ for both 
Li$_4$ and Li$_5^+$ at 50 $K$. The imaginary time correlation function 
for Li$_4$ is shown in Figure 7, averaged for atoms 1 and 2 (lower curve) ,
and for 3 and 4 (upper curve). It can be noticed that the two atoms which 
are more strongly bound (1 and 2 in Figure 2) are less delocalized than 
the other two, as may be expected.

\section{Conclusions and outlook}

\par We have introduced a novel method for simulating the statistical
mechanics of quantum nuclei interacting through first-principles potentials, 
i.e. that derive directly from the electronic structure. The scheme presented 
and discussed
here combines a path integral description of the nuclear variables with
an adiabatic, ground state, density functional description of the electronic 
degrees of freedom. In the present scheme we have choosen a specific
(NLDF-Gaussian) formulation to solve the electronic structure problem, but 
it is important to stress that any other implementation is perfectly valid 
and compatible with the present scheme, e.g. {\it ab initio} quantum 
chemical approaches like Hartree-Fock or MP2, and/or different localized 
(LCAO, LMTO) or extended (pseudopotential or augmented PW) basis sets.
Moreover, the present scheme is extremely simple to interface with any
electronic structure code, since the only input needed to compute 
the statistical weigths are the (self-consistently) converged energy and 
forces. MD sampling schemes are more involved in this respect.

We have shown the adequacy and the performance of this methodology 
by simulating quantum nuclear effects in the clusters
Li$_4$ and Li$_5^+$. The number of imaginary-time slices needed 
to achieve convergence to a relative error of 0.5\% in the nuclear 
kinetic energy is 4 at a temperature of 100 $K$, and 8 at 50 $K$. The
same level of accuracy is obtained only with 16 slices (at 100 $K$) if 
the primitive approximation to the action is used. This represents a gain
of a factor of 4, which is very important due to the high computational
cost of these simulations. The level of gain depends, however, on the 
shape of the potential energy surface sampled by the nuclei.

The results presented here for the above clusters show that, at temperatures
below 50 $K$, quantum nuclear effects are crucial to account for their
structural and electronic properties. Pair correlation functions are quite 
broadened with respect to the classical counterparts, to a level that
similar distributions would be obtained for an effective temperature of 
about 150 $K$ if only thermal, and not also quantum, fluctuations were 
considered. The results at $T=50~K$ can be considered to be representative
of the ground state, as quantum nuclear effects largely overcome thermal
motion. Electronic properties like one-electron eigenvalues and dipole 
moments show the same type of broadened distributions, although the 
effective classical temperature appears to be slightly lower, around 100 $K$, 
instead of 150 $K$. This is to emphasize that quantum effects cannot be 
readily mimicked by adding extra thermal fluctuations, because the correlations 
involved are of a completely different character: quantum motion is coherent 
while thermal motion is incoherent.

The simulation technology presented here opens up the possibility of studying 
the role of light nuclei, especially protons, in biological and chemical
systems. The advantage of describing the electronic component using a 
localized basis set (as Gaussians), as opposed to a PW basis set, for 
isolated clusters and molecules as well as for gas-phase reactions, is
obvious because the vacuum around is easily taken into account. In
addition, very often reactive condensed-phase systems and large biological 
molecules do not need a full quantum description because the relevant chemical 
processes occur in a circumscribed region of space. For example, enzymatic 
reactions require a first-principles electronic description only in the 
vicinity of the active site, while the rest of the system can be treated by 
means of classical force fields. 
Therefore, hybrid schemes that combine quantum and classical mechanical 
descriptions in different spatial regions~\cite{cpl} are appropriate for 
these situations and, in conjunction with the present methodology, 
constitute a general computational approach appropriate for studying the 
effects of nuclear delocalization in the above situations. In fact, an 
insight into the problem of quantum hydrogen-bonding in water has already 
appeared in the literature~\cite{science}, thus signalling the time for a 
new and exciting area of multidisciplinary research.

\vspace{1truecm}
{\bf Acknowledgements} C. C. would like to thank the Department of Science
and Technology, New Delhi for financial support (Grant  No. SP/S1/H-36/94). 
D. A. E. thanks Fundaci\'on Antorchas and Universidad de Buenos Aires for
financial support, and ICTP for hospitality. R. O. W. thanks Saverio Moroni 
for helping with the parallelization of the code, and J. K. thanks Dominik 
Marx for making accessible his results prior to publication and for helpful 
discussions. Useful discussions with Erio Tosatti and Pietro Ballone are 
also acknowledged. The ab initio path integral simulations have been performed 
in the IBM SP2 parallel machine at SISSA-Trieste.

\newpage

\begin{figure}
\caption{A schematic representation of a potential energy surface with two
minima separated by a barrier. Dashed lines represent different possible
values of the zero-point-energy (ZPE): (a) ZPE is larger than the barrier
(resonant regime), (b) ZPE is between the bottom of the higher well and the
top of the barrier (tunneling regime), (c) ZPE is below the bottom of the
higher well (classical regime).}
\end{figure}

\begin{figure}
\caption{The single isomer of Li$_4$ (left panel), and the four isomers
of $Li_5^+$ (right panel). Isomer IV is the most stable, followed by
isomer III. Isomers I and II are quite higher in energy, and are almost
degenerate.}
\end{figure}

\begin{figure}
\caption{Dependence of the energetics of Li clusters on the size of 
the basis set (BS). BS are ordered in increasing size.}
\end{figure}

\begin{figure}
\caption{Pair correlation functions g(r) for Li$_4$. The upper panel shows
the classical results for temperatures of 50 (solid line), 100 (dashed line),
and 200 (dotted-dashed line) $K$. The lower panel shows the quantum (dashed
line) and classical (solid line) results for $T=50~K$.}
\end{figure}

\begin{figure}
\caption{Pair correlation functions g(r) for Li$_5^+$. The upper panel shows
the classical results for temperatures of 50 (solid line), and 100 (dashed 
line) $K$. The lower panel shows the quantum (dashed line) and classical 
(solid line) results 
for $T=50~K$. The insets show the partial $g(r)$ for two groups of atoms, 
one forming
the central triangle and the other consisting of the two external atoms.}
\end{figure}

\begin{figure}
\caption{Distribution of HOMO and LUMO one-electron eigenvalues for Li$_5^+$. 
The upper panel shows the classical results for temperatures of 50 (solid 
line), and 100 (dashed line) $K$.
The lower panel shows the quantum (dashed line) and classical (solid line)
results at $T=50~K$.}
\end{figure}

\begin{figure}
\caption{Root-mean-square of the imaginary-time correlation function for 
Li$_4$ at 50 $K$ plotted versus $k$ (with $0 \le k= \beta \hbar /P \le 
\beta \hbar)$, averaged over atoms 1 and 2 (lower curve), and over atoms 
3 and 4 (upper curve). $P$ is the Trotter number, 8 in this case.}
\end{figure}

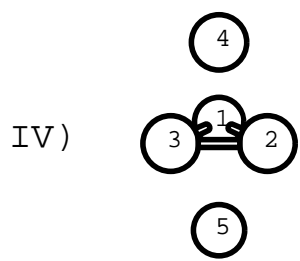
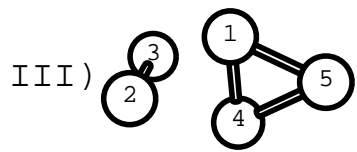
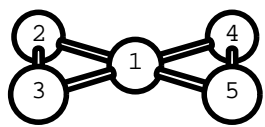
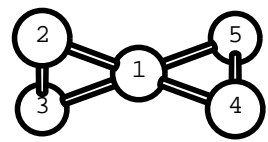
\end{document}